# Current Status of the Inner Boundaries: Short Review the Last 30 Years

**Alexander Herega and Victor Volkov**

Odessa National Academy of Food Technology, Odessa, 65039, Ukraine
aherega@gmail.com

**Abstract** Short review of experimental and theoretical researches influence of inner boundaries at properties of composite materials. Reviewing articles that are made at the last, approximately, thirty years, and which demonstrate role of inner boundaries in determining the properties of composite materials.

## 1. Introduction

It is known the inner boundaries are an attributive component of composite materials [1, 2]. Inclusion of inner boundaries into the structural parameters of the material is dictated by several reasons [3-5]. Firstly, any material is a complex system and, consequently, should have inner boundaries the existence of which is a general-system specific feature [6, 7]. Secondly, the generation of inner boundaries should be considered as a mechanism of implementation of a target function of the system, i.e. conservation of the integrity: the development of interfaces in the material is known to level the excess stress. Moreover, the inner boundaries are an inevitable consequence of the processes of self-organization of a physical body and are conditioned by the random shape of clusters and macroscopic structural units [8].

It's well understood by those at least most of the survey work on the inner boundary of the solid is not possible. The aim of this review to tell about some papers on manifold materials published in the last about 30 years, and describe the influence of inner boundaries at the properties, and results of the investigated capacities for controlling the properties of composite materials.

The authors don't even set themselves the task come close to "fully" description of the state of problems in this area of research, and ask consider this a review, as a sketch, as description of interesting them papers, as an attempt to draw attention to some of the emerging trends in the study and interpretation of the role and importance of inner boundaries.

## 2. Experimental Researches

In [9] the experimental smelting of X120 high-strength steel micro alloyed by 0.0018% mass of boron with low carbon equivalent and parameter of cracking resistance during welding had been performed. The mechanical properties tests, metallography and fractography examinations were carried out on samples made from rolled strip. The correlation between parameters of structure and mechanical properties of the steel was found. The kinetics of boron redistribution in the steel after TMP at heating and isothermal expositions in the temperature range 150-600°C was investigated using Auger Electron Spectroscopy (AES) and Secondary Ions Mass-Spectroscopy (SIMS). It was found that boron is bound into nitrides or oxides and is able to form interphase segregation. For the first time the temperature ranges of most active grain boundary segregation of impurities (B, P, S, Si, N, C, Ti) were measured, and the curves of isothermal grain boundaries enrichment kinetics for these elements were plotted for the X120 steel. The temperature dependencies of effective segregation energy and diffusion coefficient of boron in the steel were determined. The value of boron segregation process activation energy is 43.6 kJ/mol. It was shown that boron segregates from solid solution to interphase boundaries at temperatures higher than 430±20°C.

The interest to solid solutions of manganese-lanthanum perovskites is related with magnetoresistance effect observed near phase transition [10]. There are possibility to regulate physical properties through substitute ions in A- and B-sublattice [11]. Objects of investigation were lanthanum-strontium perovskites with non-stoichiometry manganese prepared by sol gel method and sintering at temperatures 550, 600, 700, 800, 900°C. The main goal of the work is definition the regularity of influence of synthesis temperature to structural features, magnetic and resistive properties of nonstoichiometric manganites obtained by the sol-gel technology.

The analyze of X-ray data indicated that all samples are single-phase and have orthorhombic structure besides sample sintering at 900°C witch has rhombohedral structure. The dependence of resistance versus temperature has semiconductor type. There is not phase transition – metal-semiconductor in all investi-gated temperature range in our pressing samples unlike of ceramic lanthanum – strontium perovskites. The magnetoresistive effects are observed at low temperature. As rule it explained by tunneling via mesoscopic boundary of particles. The value of magnetoresistance increases from 15 to 25% with increase sintering temperature to 700°C whereas more increasing sintering temperature decreases value of magneto resistance. Therefore, the sintering temperature 700°C is optimal for our pressing samples.

Authors [12] study the question extreme importance: the dependence of the material strength of the type of construction in which it incarnated. The finding that concrete strength in a structure was not correctly described by the strength of specimens moulded and stored under controlled standard conditions, led to the study of the relation between actual and potential strength of concrete.

The performance of adhesively-bonded joints under monotonic and cyclic-fatigue loading has been investigated in [13] using a fracture-mechanics approach. In particular, the influence of employing different surface pretreatments for the aluminum-alloy substrates was examined. In addition uncommon and very interesting result: a back-face strain technique has been used which revealed that crack propagation, rather than crack initiation, occupied the dominant proportion of the fatigue lifetime of the single-lap joints.

In paper [14] a method for predicting the lifetime of adhesively bonded joints and components has been investigated. The theoretical predictions using different approaches to describe the variation of with the crack length, a, and applied load in the single-lap joint, have been compared and contrasted with each other, and compared with the cyclic-fatigue behavior of the lap joints as ascertained from direct experimental measurements.

The transfer of matter and charge across interfaces between two solids is related to defect relaxation in the regions near the interface [15]. A transfer rate, which exceeds the rate of defect relaxation, may lead to degradation of the interface, causing a feedback effect for the transfer process itself. Consequently, nonlinear phenomena (dissipative structures) like periodic oscillations of the interfacial properties can occur under conditions far from equilibrium. Possible mechanisms and experimental examples are discussed.

In works [16, 17] it was established that the electrical activity of internal boundaries in silicon, as well as in silicon bicrystals produced by solid state bonding method is induced by background (oxygen, carbon, aluminium) and dopant (boron) impurities and is not result of dangling bonds formation in consequence of lattices mismatch of neighbouring grains. The complex of methods and models allowing to characterize and describe the peculiarities of formation and displaying of electrical activity of internal boundaries of different types in silicon, including induced by background and dopant impurities, is proposed.

A high density of structural defects and grain boundaries distinguishes nanocrystalline materials. Due to the small grain size, a particular defect of the grain boundary topology, the so-called triple junction takes a dominant role for grain growth and atomic transport [18]. We demonstrate by atom probe tomography that triple junctions in nanocrystalline Cu have 100-300 times higher diffusivity of Ni than standard high angle grain boundaries.

In addition, a previously unexpected systematic variation of the grain boundary width with temperature is detected. The impurity segregation layer at the grain boundaries grows from the 0.7 nm at 563 K to 2.5 nm at 643 K. This variation is clearly not controlled by simple bulk diffusion. Taking this effect into consideration, the activation energies for Ni diffusion in triple junctions and grain boundaries in Cu can be determined to be $(83 \pm 10)$ and $(120 \pm 15)$ kJ/mol, respectively. Thus, triple junctions are distinguished by considerably lower activation energy with respect to grain boundaries [18].

In [19] lanthanum nickel oxide thin films were prepared by the sol-gel method. Microstructures of the films were tailored by changing sol concentration to investigate the effect of grain boundary on the transport properties of electrons in the polycrystalline lanthanum nickel oxide films. Based on the temperature dependence of the resistivity and the magnetic field dependence of the magnetoresistance at various temperatures, the factors that dominate the transport behavior in the polycrystalline lanthanum nickel oxide films were explored in terms of weak localization and strong localization. The results show that the grain boundary has a significant influence on the transport behavior of the electrons in lanthanum nickel oxide films at a low-temperature region, which can be captured by a variable-range hopping model. The increase of metal-insulator transition temperature is ascribed to Anderson localization in grain boundary. At a high-temperature region, electron-electron scattering and electron-phonon scattering predominates in the films. In this case, the existence of more grain boundary shows a minor effect on the transport behavior of the electrons but elevates the residual resistivity of the films.

Authors [20] performed electrical transport measurements on α-sexithiophene crystalline grains using a dual-probe atomic force microscopy system having two independently controlled cantilever probes. The field-effect transistor characteristics were measured by varying the distances between the two probes brought in contact with the surface of the grains. It was clearly shown by the transfer line method that the grain boundary is the dominant factor limiting the electrical properties of organic thin films. Moreover, the hole transport across the grain boundary was found to be more affected by the oxygen hole doping than that within the crystalline grain.

## 3. Models, Conceptions, Intentions

The paper [21] considers experimental and theoretical principles of accounting for curvature in multiscale computer simulation of the behavior of inter-faces as independent planar subsystems in a loaded polycrystal. Dynamics of matter and energy flows along grain boundaries is studied using a hybrid discrete-continuous method of excitable cellular automata, which combines synergetic principles of discrete switching and laws of continuum mechanics. It is shown that these flows have a rotational-wave nature and depend on the loading conditions of a grain boundary.

In [22] analyzed of two kinds of multiple cracking in a loaded solid with a brittle coating. First is the front development of opening mode cracks, while the second is the formation of a system of sliding mode cracks along conjugate directions of maximum tangential stresses. The mesomechanics of each kind of multiple cracking is discussed. The conditions, which provide an increase in both the strength and the ductility of a loaded solid experiencing multiple cracking, are formulated; the essence of the model is the growth of cracks in the sliding mode along conjugate directions of maximum tangential stresses.

Physical mesomechanics treats a solid under loading as a multilevel system where micro-, meso- and macro levels are self-consistent [23]. The element base for the scale levels of plastic deformation of a solid is offered which allows describing any kind of plastic flow as a combination of definite elements of this element base. Special attention is paid to description of fragmentation of a solid under loading as the main mechanism of plastic flow at the meso level. Fracture is the final stage of solid fragmentation when its scale changes from meso macro level.

Interesting results of computer simulation and interpretation of field ion microscopy images of ion-irradiated platinum are presented in [24]. Formation of block nanocrystalline structures of 1–5 nm are observed in subsurface volumes not less than 20 nm from surface was observed as a result of ion irradiation. Quantitative analysis shows that block boundaries are small-angle with disorientation angle less than (6±2) degrees.

The effect of grain boundaries on the special features inherent to the evolution of atomic collision cascades and formation of radiation-affected regions in vanadium crystallites was investigated in [25]. The presence of grain boundaries in the material was found to have a considerable impact on the radiation damage pattern, specifically on the radiation-induced defect distribution and size. Grain boundaries were shown to serve as a barrier to propagation of atomic displacement cascades and to accumulate most of radiation-induced defects. Relaxed vanadium crystallites contained a relatively small number of clusters made up of point defects, i.e. vacancies and intrinsic interstitial atoms.

In [26] performed a molecular dynamics simulation of diffusion processes near the grain boundary. Calculations are carried out for the symmetric tilt boundary $S = 5$ at elevated temperatures. It is shown that near the grain boundary region high-temperature heating causes noticeable atomic displacements and thus governs active grain boundary diffusion. The calculation results demonstrate that the grain boundary diffusion parameters can be estimated with a rather high accuracy based on molecular dynamics calculations. This gives an opportunity to study the atomic mechanisms of how temperature and external mechanical fields influence diffusion processes and structural rearrangement both near the grain boundary and within the grain bulk.

The behavior of a large-angle boundary in a copper bicrystal under shear loading was studied by molecular dynamics simulation [27]. It is shown that grain boundary sliding can involve migration of the grain boundary perpendicular to the loading direction. The study covered both boundaries with a perfect symmetric structure and boundaries with a complex symmetric structure. Atomistic mechanisms of grain boundary sliding in the two cases were analyzed. It is found that the grain boundary structure affects the dynamics of the motion. Moreover, it is revealed that the grain boundary velocity depends on the value of applied load. The disclosed behavior of grain boundaries under shear deformation can exert an appreciable effect on microstructural changes of material and hence on its properties and peculiarities under loading.

The study [28] demonstrates the possibility to suppress the ductile brittle transition in bcc-structured steels at low strain temperatures on the example of pipe steel subjected to severe plastic deformation. The suppression of the ductile-brittle transition in the material is associated with structural changes in its planar subsystem (surface layers and grain boundaries in polycrystals) and substructure formation in its 3D crystalline subsystem.

The role of internal boundaries as an integral part of the structure of composite materials is discussed in [3]. The computer model of the percolation structure of the composites is proposed in this paper. The model based on the Monte-Carlo algorithms. The two and three dimensional composites model is studied and parameters of percolation clusters formed in the model are calculated.

This approach is actually creating two cluster systems – the particle and the internal boundaries, which are the background for each other, as in mosaic by M. Escher, in contrast to the "white" and "black" clusters, discussed in [29], internal boundaries forms the percolation cluster, which belongs to the known class of percolation theory problems with the zero threshold [30].

S. Trugman and A. Weinrib [30] believe their model gives a useful description of the interstitial space in rocks. In addition, they showed that in a percolation cluster system (for instance, in a semiconductor) with a certain percolation level, for particles with relevant energy it is not substantial whether the volume fraction of the conductive phase is very small. This causes association with a boat in a bay with reefs: the boat needs a layer of water above the reefs just exceeding its sea gauge in order to pass through the aquatic area.

In the model [1] it is possible to have several options of direct decrease of the system's percolation threshold. Firstly, it is variation of a control parameter of the model – the distance at which the elements are connected; this naturally decreases the power of the percolation cluster with the growth of the allowed distance. The known problems solved in such model include the definition of electrical conduction of random nets, the problems of hopping conduction, spontaneous magnetization in ferromagnetic, etc. [29].

Other options are related to the possibilities of a substantial modification of the structural elements of clusters. They are two: the first one is trivial leading to predictable results, such as substitution of circles, used in the model, on the circumference; the second is the use as structural elements or a triangle or Sierpinski carpet or Koch snowflake, i.e. the pre-fractals of a specific generation with dimensions that are dozens of times smaller than the typical sizes of the percolation field. This allows an effective variation of such parameters, as the domain connection, the correlation length, the degree of space filling, lacunarity, etc. [31].

The model of force field of the multiscale net of inner boundaries was proposed in [2]. Author assume that straight sections of the Sierpinski carpet correspond to quasilinear inner boundaries of the material, and analytically determine the force fields generated by multiscale nets of the inner boundaries of the Sierpinski pre-fractals of three types, which differ in symmetry, on the random step of partition. This allows using regular fractals, according to the principle of superposition, to calculate fields of mechanical stress, generated by the nets of inner boundaries with random configurations.

The paper [32] presents an approach to description of behavior of media and materials with a great number of inner interfaces, for example, polycrystals, nanostructured materials, media with a numerous number of fine inclusions, etc. The basic idea is to introduce a new parameter – the area of inner interfaces – and to determine the thermodynamic force conjugate to this parameter. The relationships for heat and mass fluxes and for rates of chemical transformations as well as generalized linearized equations of state for media of different types are given.

Numerous studies of dependence on the material properties from the structure usually contain photographs of the surfaces or sections, making possible to discriminate the structure of samples with differing properties. Authors of numerous articles give also verbal description of the photos, used the terms which intuitive transparent, but not have rigorous definitions, calculation algorithms, quantitative component. For example, "inhomogeneous distribution of cracks", "structured placement of filler", "complex configuration location of defects", etc.

In this connection, it is appropriate to introduce the characteristic that would allow quantifying the level of the structure order – the relative degree of the structure orderliness of the images [33, 34]. It is a quantitative characteristic, which is calculated by operable algorithm based on the concept of the information entropy.

Relative degree of orderliness is an integrality characteristic of structure, which is an alternative descriptive term such as "structured" and "heterogeneity" and allows characterizing quantitatively the drift characteristics of the material.

In [35] focuses on an application of the new and original computational methods to analyze the interfacial stress-transfer behavior and to prognosis the macromechanical properties and behavior of polymer composites in taking into consideration the real interfacial attributes and dynamics of molecular interrelation of the constituents.

The potential of applying the fractal analysis to the description of the structure and properties of polymer composites dispersedly-filled with nano- and microsize particles is discussed [36]. Much attention is paid to such important for above media problems as structure and properties of the inter-phase layers and aggregation of disperse filler particles in a polymeric composite matrix. Qualitative evaluations and physical interpretation of the alteration of the properties (reinforcement effect) of polymer composites filled with disperse particles at different scale levels is considered, taking into account the fractal parameters.

A consistent picture of the evolution of the defect structure is presented [37] for a solid undergoing plastic deformation, when defects are multiplied and arbitrary hierarchical structures are formed. The complete problem is reduced to a description of the kinetics of the formation of a new structural level of plastic deformation, a field-theoretic description of a single defect, and a statistical description of ensembles of defects. For the first time a theoretical scheme is constructed for the processes of developed plastic deformation in which a hierarchically connected multilevel structure of defects is formed. A solid with defects is treated as a highly nonequilibrium state of a crystal subjected to an intense external action.

## 4. Conclusions

In 2007 year authors [38] writing, that the current models for polycrystals address only the grain texture (i.e. the statistical parameters of crystal lattices), not the presence or character of grain boundaries, nor the size of the grains. But understanding of quantifying and predicting the role of grain size and grain boundary character on strength and ductility. This is an essential component required for the design of composites to resist stresses.

Moreover, it is necessary to agree with S. Mileiko [39]: interior boundaries determine unique properties of the composites. Awareness and understanding of this are a necessary step on the eve of a new jump of composites properties.

Really, the inner boundaries determine unique properties of the composites.

————————————————————

# References

[1] A. Herega Physical aspects of self-organization processes in composites. 1. Simulation of percolation clusters of phases and of inner boundaries. // Nanomechanics Science and Technology. – 2013. – V. 4, №2. – P. 119-132.

[2] A. Herega Physical aspects of self-organization processes in composites. 2. The structure and interaction of inner boundaries. // Nanomechanics Science and Technology. – 2013. – V. 4, №2. – P. 133-143.

[3] A. Herega et al. Percolation Model of Composites: Fraction Clusters and Internal Boundaries. // International Journal of Composite Materials. – 2012. V. 2(6). – P. 142-146.

[4] A. Herega, V. Vyrovoy Inner boundaries of composites: multiscale character of structures and properties of force fields, Proc. IV All-Russian Symp. "Mechanics of Composite Materials and Structures," Moscow: Russian Academy of Sciences, Institute of Applied Mechanics. – 2012. – V. 2. – P. 204-209.

[5] V. Vyrovoy, A. Herega, O. Korobko Oscillatory interaction of hierarchically subordinated structures of the heterogeneous material, Proc. Conf. "Modeling-2010," Kiev: G. E. Pukhov Institute of Modeling Problems in Power Engineering, National Academy of Sciences of Ukraine. – 2010. – P. 253-260.

[6] A. Uemov Systems and System Parameters. / Problems of Formal Analysis of Systems, Moscow: Vysshaya Shkola Press, 1968.

[7] L. Bertalanffy, The general system theory. A critical review, in: Studies into the General System Theory, Moscow: Progress Press. – 1969. – P. 23-82.

[8] V. Solomatov, V. Vyrovoy, A. Bobryshev The Polystructure Theory of Composite Construction Materials, Tashkent: FAN Press, 1991.

[9] Yu. Morozov, I. Simbukhov, D. Dyakonov Study of microstructure and properties of ultrahigh-strength pipe steel of strength category x120 prepared under laboratory conditions. // Metallurgist. – 2012. – V. 56 (7-8). – P. 510-518.

[10] V. Turchenko et al. Structural features, magnetic and resistive properties of nanoparticle perovskites, prepared by sol gel method. / Proceedings of IV International Conference "Functional Nanomaterials and High-purity substances". – Suzdal, 1-5 October 2012. – P. 87-88.

[11] M. Salamon, M. Jaime The physics of manganites: Structure and transport. // Rev. Mod. Phys. – 2001. – V.73. – P. 583-628.

[12] M. Soutsos, D. Breysse, V. Garnier, A. Goncalves, A. Monteiro Estimation of on-site compressive strength of concrete. / Non-Destructive Assessment of Concrete Structures: Reliability and Limits of Single and Combined Techniques. RILEM State of the Art Reports. – 2012. – V. 1. – P. 119-186.

[13] H. Hadavinia, A.J. Kinloch, M.S.G. Little, A.C. Taylor The prediction of crack growth in bonded joints under cyclic-fatigue loading I. Experimental studies. // International Journal of Adhesion and Adhesives. – 2003. – V. 23, Is. 6. – P. 449-461.

[14] H. Hadavinia, A.J. Kinloch, M.S.G. Little, A.C. Taylor The prediction of crack growth in bonded joints under cyclic-fatigue loading. II. Analytical and finite element studies. // International Journal of Adhesion and Adhesives. – 2003. – V. 23, Is. 6. – P. 463-471.

[15] J. Jurgen Oscillatory kinetics at solid-solid phase boundaries in ionic crystals. // Solid State Ionics. – 2000. – V. 131. – P. 129-142.

[16] A. Fedotov, A. Mazanik, A. Ulyashin Electrical activity of grain boundaries in silicon bicrystals and its modification by hydrogen plasma treatment // Solar Energy Materials and Solar Cells. – 2002. – Vol. 72. – P. 589-595.

[17] A. Fedotov, A. Mazanik, E. Katz et al. Electrical activity of tilt and twist grain boundaries in silicon // Solid State Phenomena. – 1999. – Vols. 67-68. – P. 15-20.

[18] M. Reda Chellali, Z. Balogh, G. Schmitz Nano-analysis of grain boundary and triple junction transport in nanocrystalline Ni/Cu. // Ultramicroscopy. – 2013. – V. 132. – P. 164-170.

[19] M.W. Zhu, Z.J. Wang, Y.N. Chen, H.L. Wang, Z.D. Zhang Effect of grain boundary on electrical properties of polycrystalline lanthanum nickel oxide thin films. // Applied Physics A. – 2013. – V. 112, Iss. 4. – P.1011-1018.

[20] M. Hirose, E. Tsunemi, K. Kobayashi, H. Yamada Influence of grain boundary on electrical properties of organic crystalline grains investigated by dual-probe atomic force microscopy. // Appl. Phys. Lett. – 2013. – V. 103. – P. 173109-112.

[21] D.D. Moiseenko, V.E. Panin, T.F. Elsukova Role of Local Curvature in Grain Boundary Sliding in a Deformed Polycrystal. // Phys. Mesomech. – 2013. – V. 16, No. 4. – P. 335-347.

[22] V.E. Panin, R.V. Goldstein, S.V. Panin Mesomechanics of multiple cracking of brittle coatings in a loaded solid. // International Journal of Fracture. – 2008. – V.150, iss. 1-2. – P. 37-53.

[23] V.E. Panin Plastic deformation and fracture of solids at the mesoscale level. // Materials Science and Engineering A. – 1997. – Vol. 234. – P. 944-948.

[24] E. Medvedeva, S. Alexandrova Computer simulation of field ion micrographs of irradiated platinum. / Proceedings of IV International Conference "Functional Nanomaterials and High-purity substances". – Suzdal, 1-5 October 2012, pp. 55-57.

[25] S.G. Psakhie, K.P. Zolnikov, D.S. Kryzhevich, A.V. Zheleznyakov, V.M. Chernov Atomic collision cascades in vanadium crystallites with grain boundaries. // Phys. Mesomech. – 2009. – V. 12, No. 1-2. – P. 20-28.

[26] S.G. Psakhie, K.P. Zolnikov, D.S. Kryzhevich Calculation of diffusion properties of grain boundaries in nanocrystalline copper. // Phys. Mesomech. – 2008. – V. 11, No. 1-2. – P. 25-28.

[27] A.I. Dmitriev, A.Yu. Nikonov, S.G. Psakhie Atomistic mechanism of grain boundary sliding with the example of a large-angle boundary $S = 5$. Molecular dy-

namics calculation. // Phys. Mesomech. – 2011. – V. 14, No. 1-2. – P. 24-28.

[28] V.E. Panin, L.S. Derevyagina, N.M. Lemeshev, A.V. Korznikov, A.V. Panin, M.S. Kazachenok On the Nature of Low-Temperature Brittleness of BCC Steels. // Phys. Mesomech. – 2014. – V. 17, No. 2. – P. 89-95.

[29] B.I. Shklovskii, A.L. Efros Electronic Properties of Doped Semiconductors. – Heidelberg: Springer, 1984. – 388 p.

[30] S. Trugman, A. Weinrib Percolation with a Threshold at Zero: a New Universality Class. // Physical Review B. – 1985. – V. 31, No. 5. – P. 2974-2980.

[31] A. Herega, V. Vyrovoy Computer modeling of the inner interfaces as elements of material structure, Proc. Conf. "Modeling-2008," Kiev: G. Pukhov Institute of Modeling Problems in Power Engineering, National Academy of Sciences of Ukraine, pp. 195–199, 2008.

[32] A. Knyazeva On simulation of irreversible processes in materials with a great number of inner interfaces. // Phys. Mesomech. – 2003. – V. 6, No. 5. – P. 11-27.

[33] A. Herega On One Criterion of the Relative Degree of Ordering in Images. // Technical Physics. – 2010. – V. 55, No. 5. – P. 741-742.

[34] A. Herega On correlation of properties and the degree of orderliness of composite structure. // Proceeding of Odessa National Academy of Food Technologies. – 2014. – V. 46 (1). – P. 236-239. (In Russian).

[35] Yu. Yanovsky, I. Obraztsov Some aspects of computer modeling of advanced polymer composite materials structure and micromechanical properties. // Phys. Mesomech. – 1998. – V. 1, No. 1. – P. 129-135.

[36] Yu. Yanovsky The mechanics of nano-structured materials and composites. From fractal description to macro scale. // Vestnik of N.I. Lobachevsky Nizhny Novgorod University. – 2011. – V. 4 (5). – P. 2641-2643. (In Russian).

[37] A. Olemskoĭ, I. Sklyar Evolution of the defect structure of a solid during plastic deformation. // Sov. Phys. Usp. – 1992. – V. 35, No. 6. – P. 455-480.

[38] E. Homer, B. Adams, R. Wagoner Recovering Grain Boundary Inclination Parameters through Oblique Double Sectioning. // Metallurgical and Materials Transactions A. – 2007. – V.38. – P. 1575-86, 2007.

[39] S. Mileiko Composites and nanostructures. // Composites and Nanostructures. – 2009. – V. 1. – P. 6-37.